\documentstyle[12pt]{article}
\input amssym.def       
\input amssym.tex       

\def\goth{\frak}          
\def\double{\Bbb}         
\def\ccal{\cal}           

\def\cc{{\double C}}     
       
\def\rr{{\double R}}     
\def\zz{{\double Z}}

\def\aa{{\cal A}}

\def\dd{{\cal D}} 
\def\gg{{\goth g}}        
\def\hh{{\cal H}}
\def\hhh{{{\double H}}}   

\def\mm{{{\ccal M}}}

\def\aa{{\cal A}}
\def\dd{{\cal D}} 
\def\hh{{\cal H}}

\def\lll{{\cal L}}

\def\t{\,{\rm tr}\,} 

\def\ddd{{\,\hbox{$\partial\!\!\!/$}}}
 
\def\dee{\,\hbox{\rm D}}
\def\de{\,\hbox{\rm d}}

\def\lb{\left[} 
\def\rb{\right]}

\def\ot{\otimes}
\def\op{\oplus}

\def\bb{\begin{eqnarray}}
\def\ee{\end{eqnarray}}
\def\eee{\nonumber\end{eqnarray}}
\def\pp{\pmatrix}
\def\qq{\quad}

\begin{document}

\hsize 17truecm
\vsize 24truecm
\font\twelve=cmbx10 at 13pt
\font\eightrm=cmr8
\baselineskip 18pt

\begin{titlepage}

\centerline{\twelve CENTRE DE PHYSIQUE THEORIQUE}
\centerline{\twelve CNRS - Luminy, Case 907}
\centerline{\twelve 13288 Marseille Cedex 9}
\vskip 3truecm

\centerline{\twelve 
On Connes' new principle of general relativity}
\centerline{\twelve Can spinors hear the forces of
spacetime?}

\bigskip

\begin{center}
{\bf Lionel CARMINATI}
\footnote{\, and Universit\'e de Provence\\
\indent carminati@cpt.univ-mrs.fr 
\qq iochum@cpt.univ-mrs.fr
\qq schucker@cpt.univ-mrs.fr}\\
{\bf Bruno IOCHUM $^{1}$ }
 \\
\bf Daniel KASTLER
\footnote{ and Universit\'e de la M\'editerran\'ee} \\
\bf Thomas SCH\"UCKER $^{1}$
\end{center}

\vskip 1 truecm
\leftskip=1cm
\rightskip=1cm
\centerline{\bf Abstract} 

\medskip

Connes has extended Einstein's principle of general
relativity to noncommutative geometry. The new
principle implies that the Dirac operator is covariant
with respect to Lorentz and internal gauge
transformations and the Dirac operator must include
Yukawa couplings. It further implies that the action
for the metric, the gauge potentials and the Higgs
scalar is coded in the spectrum of the covariant Dirac
operator. This {\it universal} action has been
computed by Chamseddine \& Connes, it is the coupled
Einstein-Hilbert and Yang-Mills-Higgs action. This
result is rederived and we discuss the physical
consequences.

\vskip 0.5truecm
PACS-92: 11.15 Gauge field theories\\ 
\indent
MSC-91: 81T13 Yang-Mills and other gauge theories 
 
\vskip 1truecm

\noindent nov. 1996
\vskip 1truecm
\noindent CPT-96/P.3413\\
\noindent hep-th/yymmxxx
 
\vskip0.1truecm

 \end{titlepage}

The dynamical variable of gravity is the metric on
spacetime. Einstein used the matrix
$g^{\mu\nu}(x)$ of the metric $g$ with respect to a
coordinate system $x^\mu$ to parameterize the set of
all metrics on a fixed spacetime $M$. The coordinate
system being unphysical, Einstein required his field
equations for the metric to be covariant under 
coordinate transformations, the principle of general
relativity. Elie Cartan used tedrads, {\it rep\`eres
mobiles}, to parameterize the set of all metrics. This
parameterization allows to generalize the Dirac
operator $\dd$ to curved spacetimes and also
reformulates general relativity as a gauge theory
under the Lorentz group. Connes
\cite{congr} goes one step further by relating the set
of all metrics to the set of all Dirac operators. The
Einstein-Hilbert action, from this point of view, is the
Wodzicki residue of the second inverse power of the
Dirac operator
\cite{wod} and is computed most conveniently from
the second coefficient of the heat kernel expansion of
the Dirac operator squared. The heat kernel expansion
\cite{heat}
is an old friend \cite{friend} from quantum field
theory in curved spacetime, from its formal relation to
the one-loop effective action 
\bb S_{eff}=\t \log (\dd/\Lambda),\qquad \Lambda\
{\rm a\ cut\ off}.\ee 
This relation has been used by
Sakharov \cite{sak} to induce gravity from quantum
fluctuations, leading however to a negative Newton
constant \cite{sakcor}. 

By generalizing the metric, the Dirac operator plays a
fundamental role in noncommutative geometry. To
describe Yang-Mills theories, Connes considers
the product of spacetime and internal space in this new
geometry, a natural point of view because the
fermionic mass matrix qualifies as Dirac operator on
internal space. Now comes the first miracle: in the
same sense that Minkowskian geometry forces the
electric field to be accompanied by a magnetic field,
noncommutative geometry forces certain Yang-Mills
fields to be accompanied by a symmetry breaking
Higgs field \cite{ssb}. This miracle takes
place only in a tiny class of Yang-Mills-Dirac theories
\cite{tiny}
and the second miracle is that the standard model of
electro-weak and strong forces is in this class. 

Now we are ready for the third miracle. It extends
Einstein's principle of general relativity to 
noncommutative geometry. To generalize the Dirac
operator from flat to curved spacetime (locally), it is 
sufficient to write the Dirac operator first in flat
spacetime but with respect to noninertial coordinates.
A straightforward calculation produces the {\it
covariant} Dirac operator that contains the spin
connection $\omega$. Although of vanishing
curvature, $\omega$ contains a lot of physics, e.g. the
centrifugal and Coriolis accelerations in the
coordinates of the rotating disk, the quantum
interference pattern of neutrons \cite{bonse} in
oscillating coordinates. Then, the generalization to
curved space is easy where $\omega$ describes the
(minimal) coupling of the spinor to the gravitational
field. From this point of view, the covariant Dirac
operator is obtained by acting with the
diffeomorphism group on the flat Dirac operator. But
the diffeomorghism group is just the automorphism
group of the associative (and commutative) algebra
${\cal C}^\infty(M)$ representing spacetime in the
new geometry. On the other, hand the product of
spacetime and internal space is represented in this
geometry by the tensor product of ${\cal C}^\infty(M)$
with a matrix algebra. Its automorphism group is the
semi-direct product of the diffeomorphisms and the
group of gauge transformations, the diffeomorphisms
are the outer, the gauge transformations are the inner
automorphisms. And what do we get when this entire
automorphism group acts on the flat Dirac operator?
We get the completely covariant Dirac operator
containing the spin connection, the gauge connection
and the Higgs \cite{tresch}. In other words, we get the
minimal couplings of the Dirac spinor to the
gravitational and Yang-Mills fields and its Yukawa
couplings to the Higgs field. In Connes' words, the
Higgs and Yang-Mills fields are noncommutative
fluctuations of the metric. (Abelian Yang-Mills
theories do not have such fluctuations.) Accordingly,
Connes generalizes Einstein's principle of general
relativity by postulating that only the intrinsic
properties of the covariant Dirac operator be relevant
for physics. Here intrinsic means invariant under
automorphisms. Thus, these properties must concern
 the spectrum only. 

So far we have only the kinematics of the metric (and
its fluctuations). To get its dynamics, Einstein
developed the full power of the principle of general
relativity and derived the Einstein-Hilbert action. This
is what Chamseddine \& Connes \cite{cc} now did also
for the fluctuations of the metric, the Yang-Mills and
Higgs fields. The fourth miracle is that this action
comes out to be the Einstein-Hilbert action
accompanied by
the Yang-Mills action, by the covariant
Klein-Gordon action and by the symmetry breaking
Higgs potential. 

In even dimensions, the spectrum of the Dirac
operator is even and it is sufficient to
consider the positive part of the spectrum which in
the  Euclidean is conveniently characterized by its
distribution function
$S_\Lambda$ equal to the number of eigenvalues
smaller than the positive real $\Lambda$,
\bb S_\Lambda=\t f(\dd/\Lambda).\ee
Here $f(u)$ denotes the characteristic function of the
unit interval. Instead, if $f$ was the logarithm, this
trace, after a proper renormalization, would be
Sakharov's induced gravity action.

\section{The Dirac operator}

The starting point of Chamseddine \& Connes' action
calculation \cite{cc} is the covariant Dirac operator
$\dd$ of the standard model. $\dd$ acts on a multiplet of
Weyl spinors. These are therefore dynamical fields
coupled minimally to fixed, adynamical fields, the
gravitational, the Yang-Mills and the Higgs fields. 
\bb\dd=\pp{
i[\ddd\ot 1_L\,+\,
{e^\mu}_j\gamma^j\ot\rho_L(A_\mu)]&\gamma_5
\ot\Phi
\cr \gamma_5\ot\Phi^*&
i[\ddd\ot 1_R\,+\,
{e^\mu}_j\gamma^j\ot\rho_R(A_\mu)]},\ee
with
\bb\ddd={e^\mu}_j\gamma^j\left(\frac{\partial}
{\partial x^\mu}\,+\,{\textstyle\frac{1}{4}}\,
\omega_{ab\mu}\gamma^{ab}\right),\ee
${e^\mu}_j(x)\partial/\partial x^\mu$ is an
orthonormal frame of the tangent bundle of an
compact, Euclidean spacetime
$M$,
$\omega$ the torsionless spin connection with respect
to this frame, $A$ is the Yang-Mills field, the
connection of the internal Lie algebra $\gg$,
$\rho_L$ is the unitary representation of $\gg$ on the
Hilbert space
$\hh_L$ of left-handed spinors, $1_L$ is the identity
on $\hh_L$, likewise for the right-handed spinors,
$\cdot_R$.
$\Phi$ is the scalar multiplet. It must necessarily be a
subrepresentation of $\hh_L^*\ot\hh_R\,\op\,
\hh_L\ot\hh_R^*$. 
 Our conventions for Dirac matrices are:
\bb \gamma^a\gamma^b+\gamma^b\gamma^a=
2\eta^{ab}1_4,&\gamma^{ab}={\textstyle\frac{1}{2}}
[\gamma^a,\gamma^b],&\gamma^{\mu
*}=\gamma^\mu, \cr \eta^{ab}={\rm
diag}(+1,+1,+1,+1),
&\gamma_5^2=1_4,&
\gamma^{ *}_5=\gamma_5.\ee

Let us spell out the representations for the standard
model. The group is $SU(2)\times U(1)\times SU(3)$,
\bb\hh_L&= &\left(\cc^2\ot\cc^N\ot\cc^3\right)\
\op\ 
\left(\cc^2\ot\cc^N\right),\\
\hh_R&=&\left((\cc\op\cc)\ot\cc^N\ot\cc^3\right)\ 
\op\ \left(\cc\ot\cc^N\right).\ee
The first factor denotes weak isospin, the second $N$
generations, $N=3$, and the third denotes colour
triplets and singlets. We take as basis
\bb \pp{u\cr d}_L,\ \pp{c\cr s}_L,\ \pp{t\cr b}_L,\ 
\pp{\nu_e\cr e}_L,\ \pp{\nu_\mu\cr\mu}_L,\ 
\pp{\nu_\tau\cr\tau}_L\label{leftbasis}\ee
for $\hh_L$ which is 24 dimensional and 
\bb\matrix{u_R,\cr d_R,}\qq \matrix{c_R,\cr s_R,}\qq
\matrix{t_R,\cr b_R,}\qq  e_R,\qq \mu_R,\qq 
\tau_R\label{rightbasis}\ee 
 for $\hh_R$ of 21 dimensions. The fermionic mass
matrix is
\bb\mm=\pp{
\pp{M_u&0\cr 0&M_d}\ot 1_3&0\cr
0&\pp{0\cr M_e}}\ee
with
\bb M_u:=\pp{
m_u&0&0\cr
0&m_c&0\cr
0&0&m_t},\qq M_d:= C_{KM}\pp{
m_d&0&0\cr
0&m_s&0\cr
0&0&m_b},\qq M_e:=\pp{
m_e&0&0\cr
0&m_\mu&0\cr
0&0&m_\tau}\ee
where $C_{KM}$ denotes the 
Cabbibo-Kobayashi-Maskawa matrix. Let
$\varphi=(\varphi_1,\varphi_2)^T$ be the complex
scalar doublet, its embedding in
$\hh_L^*\ot\hh_R\,\op\,
\hh_L\ot\hh_R^*$ is given by
\bb \Phi=\frac{1}{v}\,\pp{
\pp{\varphi_1M_u&-\bar\varphi_2M_d\cr 
         \varphi_2M_u&\bar\varphi_1M_d}\ot 1_3&0\cr 
0&\pp{-\bar\varphi_2M_e\cr \bar\varphi_1M_e}},
\label{emb}\ee
with $v$ denoting the vacuum expectation value.
Finally our conventions for hypercharges are:
\bb y(u_L)=y(d_L)={\textstyle\frac{1}{6}},&
y(u_R)={\textstyle\frac{2}{3}},&
y(d_R)=-{\textstyle\frac{1}{3}},\cr 
y(\nu_L)=y(e_L)=-{\textstyle\frac{1}{2}},&&
y(e_R)=-1,\cr
 y(\varphi_1)=y(\varphi_2)=-{\textstyle\frac{1}{2}}.&
&\ee

\section{The Dirac operator squared}
 
 Since the trace of the characteristic function
only counts eigenvalues, we have 
\bb S_\Lambda:=\t f(\dd/\Lambda)
=\t f((\dd/\Lambda)^2).\ee
Although the Dirac operator is only a first order
differential operator, the computation of its square is
either very long or subtle. The result is the well
known Lichn\'erowicz formula \cite{at}. In our case it
yields:
\bb \dd^2=-\Delta+E,\ee
and its trace can be computed using the heat kernel
technique \cite{heat}.
  $\Delta$ is the covariant Laplace operator
\bb \Delta&=&g^{\mu\tilde\nu}\left[\left(
\frac{\partial}{\partial x^\mu} 1_4\ot 1_\hh+
{\textstyle\frac{1}{4}}\omega_{ab\mu}\gamma^{ab}
\ot 1_\hh+1_4\ot \rho(A_\mu)\right)
{\delta^\nu}_{\tilde\nu}-{\Gamma^\nu}_{\tilde
\nu\mu}1_4\ot1_{\hh}\right]
\cr &&\qq\qq\qq\times\left[
\frac{\partial}{\partial x^\nu} 1_4\ot 1_\hh+
{\textstyle\frac{1}{4}}\omega_{ab\nu}\gamma^{ab}
\ot 1_\hh+1_4\ot \rho(A_\nu)\right]\ee
with the fermionic representation
$\rho:=\rho_L\op\rho_R$ on $\hh:=\hh_L\op\hh_R$
and $\Gamma$ are the Christoffel symbols of the spin
connection. 
$E$, for endomorphism, is a zero order operator, that is
a matrix of size
$4\dim\hh$ whose entries are functions
constructed from the adynamical bosonic fields and
their first and second derivatives,
\bb E={\textstyle\frac{1}{2}}\left[e^\mu_ce^\nu_d
\gamma^{cd}\ot1_\hh\right]\rr_{\mu\nu}\,+\,
\pp{
1_4\ot \Phi\Phi^*&-i\gamma_5\gamma^\mu
\dee_\mu\Phi\cr  
-i\gamma_5\gamma^\mu (\dee_\mu\Phi)^*&
1_4\ot\Phi^*\Phi}\label{E}.\ee
$\rr$ is the total curvature, a 2-form with values in
the (Lorentz $\op$ internal) Lie algebra represented
on (spinors $\ot\ \hh$). It contains the curvature
2-form
$R=\de\omega+{\textstyle\frac{1}{2}}[\omega,\omega]
$ and the field strength 2-form $F=\de
A+{\textstyle\frac{1}{2}}[A,A]$, in components
\bb\rr_{\mu\nu}={\textstyle\frac{1}{4}}
R_{ab\mu\nu}\gamma^{ab}\ot 1_\hh+
1_4\ot\rho(F_{\mu\nu}).\ee 
An easy calculation shows that the first term in
equation (\ref{E}) produces
the curvature scalar that we also (!)
denote by $R$,
\bb {\textstyle\frac{1}{2}}\left[e^\mu_ce^\nu_d
\gamma^{cd}\right]
{\textstyle\frac{1}{4}}R_{ab\mu\nu}\gamma^{ab}
= {\textstyle\frac{1}{4}}R1_4.\ee
 In our conventions, the curvature scalar
is positive on spheres.
 Finally $\dee$ is the covariant derivative appropriate
to the representation of the scalars.
 For more details on our conventions the reader
is referred to
\cite{gs}. Note that the decomposition
$\dd^2=-\Delta+E$ is simple thanks to the presence of
$\gamma_5$ in $\dd$. This $\gamma_5$ is deeply
rooted in noncommutative geometry.

We close this section with a remark
on the powers of $\Phi$. Here we need two, later we
will also meet four powers. In general, they are
cumbersome to compute. For the standard model, there
is a trick that comes from its noncommutative
formulation. Let us denote by $X\in su(2)$ an element
of weak isospin and by $\rho_{Lw}$ its representation
on $\hh_L$. This representation can be extended to a
representation of the quaternions $\hhh$ as
involution algebra. $\hhh$ contains $su(2)$ as the Lie
algebra of its group of unitaries. A quaternion $\phi$
is parameterized by two complex numbers, $\varphi_1$
and $\varphi_2$,
\bb \phi=\pp{\varphi_1&-\bar\varphi_2\cr 
\varphi_2&\bar\varphi_1}\in\hhh.\ee
Then, the embedding (\ref{emb}) of the scalar doublet
$\varphi$  in $\hh_L^*\ot\hh_R\,\op\,
\hh_L\ot\hh_R^*$, which is nothing but the
Yukawa couplings, takes the form of a matrix product,
\bb \Phi=\rho_{Lw}(\phi)\mm/v,\ee
and the higher powers of $\Phi$ follow easily from the
identity
\bb \phi^*\phi=\phi\phi^*=
(|\varphi_1|^2+|\varphi_2|^2)1_2=|\varphi|^21_2.\ee

\section{The trace}

Asymptotically, for large $\Lambda$, the distribution
function of the spectrum is given in terms of the
heat kernel expansion \cite{heat}:
\bb S_\Lambda=\t f(\dd^2/\Lambda^2)=
\frac{1}{16\pi^2}\,\int_M[\Lambda^4f_0a_0+
\Lambda^2f_2a_2+f_4a_4+\Lambda^{-2}f_6a_6+...]
\sqrt{\det g}\,\de^4x, \label{master}\ee
with 
\bb f_0\ =&\int_0^\infty uf(u)\de u& 
=\ {\textstyle\frac{1}{2}},\\
f_2\ =&\int_0^\infty f(u)\de u&=\ 1,\\
f_4\ =&f(0)&=\ 1,\\
f_6\ =&f'(0)&=\ 0,\\
f_8\ =&f''(0)&=\ 0,...\ee
The $a_j$ are the coefficients of the heat kernel
expansion of the Dirac operator squared,
\bb a_0&=&\t (1_4\ot1_\hh),\\ 
a_2&=&{\textstyle\frac{1}{6}}R\t (1_4\ot1_\hh)-\t E,\\
a_4&=&{\textstyle\frac{1}{72}}R^2\t (1_4\ot1_\hh)-
{\textstyle\frac{1}{180}}R_{\mu\nu}R^{\mu\nu}
\t (1_4\ot1_\hh)+
{\textstyle\frac{1}{180}}R_{\mu\nu\rho\sigma}
R^{\mu\nu\rho\sigma}\t (1_4\ot1_\hh)\cr &&+
{\textstyle\frac{1}{12}}\t (\rr_{\mu\nu}
\rr^{\mu\nu}) 
-{\textstyle\frac{1}{6}}R\t E+
{\textstyle\frac{1}{2}}\t E^2 + {\rm surface\ terms}.\ee
Let us first check the normalization $16\pi^2$ of
equation (\ref{master}). We take $M$ to be the flat
4-torus with unit radii, $\hh_L=\cc$, $\hh_R=0$ and
$A=\varphi=0$. Denote by $\psi_B$, $B=1, 2, 3, 4$,
the four components of the spinor. The Dirac
operator is
\bb \ddd = \pmatrix{
i{\partial/{\partial x^0}}&0
&-{\partial/{\partial x^3}}&
-{\partial/{\partial x^1}}+
i{\partial/{\partial x^2}}\cr
0&i{\partial/{\partial x^0}}&
-{\partial/{\partial x^1}}-
i{\partial/{\partial x^2}}&
{\partial/{\partial x^3}}\cr
{\partial/{\partial x^3}}&
{\partial/{\partial x^1}}-
i{\partial/{\partial x^2}}&
-i{\partial/{\partial x^0}}&0\cr
{\partial/{\partial x^1}}+
i{\partial/{\partial x^2}}&
-{\partial/{\partial x^3}}&0&
-i{\partial/{\partial x^0}} }
.\ee
After a Fourier transform
\bb \psi_B(x)\ =:\ \sum_{j_0,...,j_3\in\zz}
\hat\psi
_B(j_0,...,j_3)\exp(-ij_\mu x^\mu),\quad B=1,2,3,4\ee
the eigenvalue equation $\ddd\psi=\lambda\psi$ reads
\bb \pmatrix{
j_0&0&ij_3&ij_1+j_2\cr
0&j_0&ij_1-j_2&-ij_3\cr
-ij_3&-ij_1-j_2&-j_0&0\cr
-ij_1+j_2&ij_3&0&-j_0}
\pmatrix{\hat\psi_1\cr \hat\psi_2\cr \hat\psi_3\cr
\hat\psi_4}\ =\
\lambda
\pmatrix{\hat\psi_1\cr \hat\psi_2\cr \hat\psi_3\cr
\hat\psi_4}. \ee
Its characteristic equation is
$ \lb \lambda^2-(j_0^2+j_1^2+j_2^2+j_3^2)^2\rb^2=0$
and for fixed $j_\mu$, each eigenvalue
$ \lambda=\pm\sqrt{j_0^2+j_1^2+j_2^2+j_3^2}$
has multiplicity two. Therefore asymptotically for large
$\Lambda$ there are 
$ 4B_4\Lambda^4$ eigenvalues (counted with their
multiplicity) whose absolute values are smaller than 
$\lambda$, $\lambda>0$. 
$ B_4=\pi^2/2$
denotes the volume of the unit ball
in $\rr^4$ and
\bb S_\Lambda=
4{\textstyle\frac{1}{2}}\pi^2\Lambda^4
=\frac{1}{16\pi^2}\,\Lambda^4{\textstyle\frac{1}{2}}4
(2\pi)^4.\ee 

The computation of the Chamseddine-Connes action
$S_\Lambda$ for the Dirac operator of the standard
model is straightforward. We give a few intermediate
steps, a full account can be found in \cite{rom}. 
\bb a_0&=&4\dim\hh,\\
\t E&=&\dim\hh\, R + 8 \t\Phi^*\Phi
=\dim\hh\,R+8 L|\varphi/v|^2,\\
L&:=&3\t (M_u^*M_u)+3\t (M_d^*M_d)+\t
(M_e^*M_e)\cr 
&=&3(m^2_t+m^2_c+m^2_u+m^2_b+m^2_s+m^2_d)+
m^2_\tau+m^2_\mu+m^2_e,\\
a_2&=&{\textstyle\frac{4}{6}}\dim\hh\, R-\dim\hh\, R
-8 L|\varphi/v|^2= -{\textstyle\frac{1}{3}}\dim\hh\, R
-8 L|\varphi/v|^2,\\
\t \gamma^{ab}\gamma^{cd}&=&
4\left[\eta^{ad}\eta^{bc}-\eta^{ac}\eta^{bd}\right],\\
\t \rr_{\mu\nu}\rr^{\mu\nu}&=&
-{\textstyle\frac{1}{2}}\dim\hh\,
R_{\mu\nu\rho\sigma}R^{\mu\nu\rho\sigma}
-4 \t \rho(F_{\mu\nu})^*\rho(F^{\mu\nu}),\\
\t E^2&=&{\textstyle\frac{1}{4}}\dim\hh\, R^2+
2\t \rho(F_{\mu\nu})^*\rho(F^{\mu\nu})\cr &&+
8L_2|\varphi/v|^4+
8L(\dee_\mu\varphi/v)^*(\dee^\mu\varphi/v)
+4L|\varphi/v|^2R,\\ 
L_2&:=&3\t \left[M_u^*M_u\right]^2+3\t
\left[M_d^*M_d\right]^2+\t
\left[M_e^*M_e\right]^2\cr 
&=&3(m^4_t+m^4_c+m^4_u+
m^4_b+m^4_s+m^4_d)+
m^4_\tau+m^4_\mu+m^4_e.
\ee 
Using the Weyl tensor,
\bb C_{\mu\nu\rho\sigma}:=
R_{\mu\nu\rho\sigma}
-{\textstyle\frac{1}{2}}(
g_{\mu\rho}R_{\nu\sigma}-
g_{\mu\sigma}R_{\nu\rho}+
g_{\nu\sigma}R_{\mu\rho}-
g_{\nu\rho}R_{\mu\sigma})
+{\textstyle\frac{1}{6}}
(g_{\mu\rho}g_{\nu\sigma}-
g_{\mu\sigma}g_{\nu\rho})R 
,\ee
we can assemble all higher derivative gravity terms 
in $a_4$ to form the square of the Weyl tensor
\bb
C_{\mu\nu\rho\sigma}C^{\mu\nu\rho\sigma}=
R_{\mu\nu\rho\sigma}R^{\mu\nu\rho\sigma}-2
R_{\mu\nu}R^{\mu\nu}+{\textstyle\frac{1}{3}}R^2 
=2R_{\mu\nu}R^{\mu\nu}
-{\textstyle\frac{2}{3}}R^2+
{\rm surface\ term},\ee
because
$ R_{\mu\nu\rho\sigma}R^{\mu\nu\rho\sigma}
-4R_{\mu\nu}R^{\mu\nu}+R^2$
is proportional to the Euler characteristic of $M$. Then,
up to this surface term, we have
\bb -{\textstyle\frac{1}{360}} \dim\hh\left[
7R_{\mu\nu\rho\sigma}R^{\mu\nu\rho\sigma}+
8R_{\mu\nu}R^{\mu\nu}-5R^2\right]=
-{\textstyle\frac{1}{20}}\dim \hh
C_{\mu\nu\rho\sigma}C^{\mu\nu\rho\sigma}.\ee
Finally we have up to surface terms,
\bb a_4 &=& -{\textstyle\frac{1}{20}}\dim \hh\,
C_{\mu\nu\rho\sigma}C^{\mu\nu\rho\sigma}+
{\textstyle\frac{2}{3}}
\t \rho(F_{\mu\nu})^*\rho(F^{\mu\nu})\cr &&+
4L_2|\varphi/v|^4+
4L(\dee_\mu\varphi/v)^*(\dee^\mu\varphi/v)
+{\textstyle\frac{2}{3}}
L|\varphi/v|^2R.\ee

\section{The bare action}

The Chamseddine-Connes action $S_\Lambda$ is seen to
be the combined Einstein-Hilbert and
Yang-Mills-Higgs actions of the standard model plus a
higher derivative gravity term plus the conformal
scalar-gravity coupling. By normalizing the Higgs
field $\varphi$ and the Yang-Mills fields
$A^{(j)}_\mu$, $j=3,2,1$
for $su(3),\ su(2)$ and $u(1)$,  we rewrite the
Lagrangian in its conventional Euclidean form,
\bb \lll_\Lambda&=& -m^{'2}_P/(16\pi)\,R
\,+\,\Lambda'_C\cr &&+\,
1/(2g_3^2)\, \t F_{\mu\nu}^{(3)}F^{(3)\mu\nu}\,+\,
1/(2g_2^2)\, \t F_{\mu\nu}^{(2)}F^{(2)\mu\nu}\,+\,
1/(4g_1^2)\,  F_{\mu\nu}^{(1)}F^{(1)\mu\nu}\cr &&
+\,{\textstyle\frac{1}{2}}\,(\dee_\mu\varphi)^*
\dee^\mu\varphi\,+\,\lambda |\varphi|^4
\,-\,{\textstyle\frac{1}{2}}\,\mu^2|\varphi|^2
\cr &&-\,a
C_{\mu\nu\rho\sigma}C^{\mu\nu\rho\sigma}\,+\,
{\textstyle\frac{1}{12}}\,|\varphi|^2
R.\label{lagrangian}\ee 
Note the correct sign of the
following terms:  Newton's constant $m_P^2/(16\pi)$ is
positive, the three gauge couplings $g_3,\ g_2,\ g_1$
are real (positive), the kinetic term of the Higgs field
is positive, the Higgs self coupling $\lambda$ is
positive, and the symmetry breaking term $\mu$ is
real (positive). For the standard model with $N=3$
generations and $\dim\hh=45$ we get,
\bb g_3^2&=&g_2^2\ =\ 3\pi^2/N\ =\
\pi^2,\label{g23}
\\ g_1^2&=&9\pi^2/(5N)\ =\
{\textstyle\frac{3}{5}}\,\pi^2,\label{g1}\\
\lambda&=&\pi^2L_2/L^2 \ 
=\ \pi^2(1-2m_b^2/m_t^2)\,+\,O(m_\tau^2/m_t^2),
\label{stiffl}\\
\mu^2&=& 2\Lambda^2\label{stiffm},\\
a&=&\dim\hh /(320\pi^2)\ =\ 9/(64\pi^2).\ee
Before identifying Newton's constant and the
cosmological constant $\Lambda_C$ we have to shift
the Higgs field by its vacuum expectation value,
$|\varphi|=v=\mu/(2\sqrt\lambda)$. This shift
changes $m'_P$ and $\Lambda_C $ into
\bb m_P^2&=&{\textstyle\frac{1}{3\pi}}(\dim\hh-2
L^2/L_2)\Lambda^2\ =\
{\textstyle\frac{43}{3\pi}}
\Lambda^2\,+\,O(m_b^2/m_t^2)\Lambda^2,\\
\Lambda_C&=&{\textstyle\frac{1}{8\pi^2}}(\dim\hh-2
L^2/L_2)\Lambda^4\ =\
{\textstyle\frac{43}{8\pi^2}}
\Lambda^4\,+\,O(m_b^2/m_t^2)\Lambda^4.\label{stiffend}\ee

\section{The soft action}

Equations (\ref{g23},\ref{g1}) tell us $g_3=g_2$ and
$\sin^2\theta_w={\textstyle\frac{3}{8}}$, like in
$SU(5)$ grand unification. Naturally, we interpret
these relations to hold at very high energy $10^{15}$
GeV. But this means that we have to swallow two 
assumptions. The first is the big desert, from $10^3$
GeV up to $10^{15}$ GeV, the standard model remains
valid without modifications, in particular no new
particles. The second is that {\it perturbative}
quantum field theory gets through the big desert
without collapsing. Note that these two assumptions
imply \cite{cabb} that the Higgs mass is
constrained to an interval, 160 GeV $<\ m_H\  <$ 200
GeV for the gauge couplings as mesured in1979.
These assumptions also imply
\cite{bin} that the above relations evolve down to
$g_3=1.8$ and
$\sin^2\theta_w=0.21$ at 5 GeV. In the beginning of the
eighties, experiments were compatible with these
values and grand unification was en vogue. Today
with $\sin^2\theta_w=0.2315\pm 0.0005\ $
experimentally, ${\textstyle\frac{3}{8}}$ is less
attractive. Moreover $g_3=\pi$ from equation
(\ref{g23}) excludes perturbation theory. Accordingly,
we don't attach much attention to the
numerical values of the coefficients in the
Lagrangian, but we do acknowledge that we get all
necessary terms and with coefficients of the right sign.
This attitude is supported by the observation that the
Chamseddine-Connes action
$ S_\Lambda=\t f(\dd^2/\Lambda^2)$ is universal
with respect to the choice of the characteristic
function of the unit interval 
$f$. Indeed for any positive, sufficiently regular
function
$f$, we get the desired Lagrangian with correct signs
and involving --- for large $\Lambda$ --- three
additional, arbitrary positive constants $f_0,\ f_2,\
f_4$. Note that $f=\log$ of induced gravity is neither
positive nor sufficiently regular. Allowing
$f_4\not=1$ solves the problem $g_3=\pi$, keeping the
grand unification flavor  $g_3=g_2$ and
$\sin^2\theta_w={\textstyle\frac{3}{8}}$.
This is no surprise, the
Yang-Mills action with its coupling constants is still
obtained from one trace over all fermions.
At this point, we emphasize
that taking due account of the mass matrix, the
fermionic Hilbert space of the standard model has four
irreducible pieces, the quarks and the three lepton
generations. Accordingly, we will soften
$S_\Lambda$ further by taking independent traces in
each irreducible piece. This amounts to
 introducing the
`noncommutative coupling' $z$, a positive operator on
the Hilbert space, that commutes with the
representation, with the Dirac operator and with the
chirality. For the standard model, $z$ is a constant
matrix involving four positive numbers $x,\ y_1,\
y_2,\ y_3$. With respect to the basis
(\ref{leftbasis},\ref{rightbasis}), $z$ takes the form
\bb z=1_4\ot\pp{
x/3\,1_2\ot 1_N\ot 1_3&0&0&0\cr  0&1_2\ot y&0&0\cr 
0&0&x/3\,1_2\ot 1_N\ot 1_3&0\cr 
0&0&0&y},
\qq y:=\pp{ 
y_1&0&0\cr 
0&y_2&0\cr 
0&0&y_N},\ee
and the softened Chamseddine-Connes action is
\bb S_\Lambda=\t f(z\dd^2/\Lambda^2).\ee
We repeat that the form of the Lagrangian
(\ref{lagrangian}) remains unchanged and its
couplings now read:
\bb
g_3^{-2}&=&{\textstyle\frac{1}{9\pi^2}}f_4Nx,
\label{softb}\\ 
g_2^{-2}&=&{\textstyle\frac{1}{12\pi^2}}f_4(Nx+\t
y),\\  g_1^{-2}&=&{\textstyle\frac{1}{12\pi^2}}f_4(
{\textstyle\frac{11}{9}}Nx+3\t y),\\ 
\lambda&=&\frac{\pi^2L_2}{f_4L^2}\,\label{softlam}
\\  &&L\ =\ x\t M_u^*M_u+x\t M_d^*M_d+\t y
M_e^*M_e\cr  &&\qq\qq =\ 
x(m^2_t+m^2_c+m^2_u+m^2_b+m^2_s+m^2_d)+
y_3m^2_\tau+y_2m^2_\mu+y_1m^2_e,\\
&&L_2\ =\ x\t \left[M_u^*M_u\right]^2+x\t
\left[M_d^*M_d\right]^2+\t y
\left[M_e^*M_e\right]^2\cr 
&&\qq\qq=\  x(m^4_t+m^4_c+m^4_u+
m^4_b+m^4_s+m^4_d)+
y_3m^4_\tau+y_2m^4_\mu+y_1m^4_e,\\
\mu^2&=& 2\,\frac{f_2}{f_4}\,\Lambda^2,\\
m_P^2&=&{\textstyle\frac{1}{3\pi}}f_2\left[(4Nx+3\t y)
-2\,\frac{L^2}{L_2}\,\right]
\Lambda^2,\\
\Lambda_C&=&{\textstyle\frac{1}{4\pi^2}}
\left[f_0(4Nx+3\t y)
-\,\frac{f_2^2}{f_4}\,\frac{L^2}{L_2}\,\right]
\Lambda^4,\\
a&=&f_4\,\frac{4Nx+3\t y}{320\pi^2}.
\label{softe}\ee
We recover the stiff values, equations
(\ref{g23}-\ref{stiffend}), by taking for
$f$ the characteristic function of the unit
interval: $f_0=1/2,\ f_2=1,\ 
 f_4=1$, and by taking for $z$ the identity: $x=3,\
y=1_N$. The soft values avoid the problems of non
perturbative gauge couplings and a huge cosmological
constant. They can still not be interpreted at low
energies because the weak angle, 
\bb\sin^2\theta_w=\,\frac{g_2^{-2}}
{g_1^{-2}+g_2^{-2}}\,=\,\frac{Nx+\t
y}{{\textstyle\frac{20}{9}}Nx+4\t y}\,,\ee
is constrained by $0.25<\sin^2\theta_w<0.45$ and even
in the soft version, we have to swallow the big desert.
This is the subject of the next section.

\section{Renormalization flow}

In this section, we combine the soft constraints on the 
couplings (\ref{softb}-\ref{softe}) coming from
noncommutative geometry with the renormalization
flow coming from the short distance divergence of
quantum corrections. Noncommutative geometry does
to spacetime what quantum mechanics does to
phase space: it introduces an uncertainty principle.
Below the scale $\hbar$, points of phase space are not
resolved. Below the scale $1/\Lambda$, points of
spacetime are not resolved and noncommutative
geometry scoffs at short distance divergences.
We adopt the philosophy that the constraints on the
couplings are valid at the momentum cut off 
$\Lambda$. Below this energy scale, we trust in the
continuum approximation and the couplings should
run according to Wilson. 

We consider the energy dependence of the couplings
$g(t),\ t:=\log E/\Lambda$ perturbatively in the one
loop approximation and neglect threshold effects. We
also work in flat spacetime only: although the
gravitational part of the Chamseddine-Connes action
includes a curvature square term it is
non-renormalizable
\cite{stelle} and it contains negative modes \cite{tam}.
With these simplifications, the evolution of the gauge
couplings decouples and is simply logarithmic:
\bb \,\frac{\de g}{\de t}\ =:\ \beta_g,\qq 
\beta_g\ =\ {\textstyle\frac{1}{16\pi^2}}c_gg^3,\qq
4\pi g^{-2}(t)\ =\ 4\pi g^{-2}(0)-
{\textstyle\frac{c_g}{2\pi}}t.\ee
For the standard model with $N=3$ generations, the
$\beta$ functions of the three gauge couplings are
in this approximation given by \cite{jones}
\bb c_3=-7,\qq c_2=-{\textstyle\frac{19}{6}},
\qq c_1={\textstyle\frac{41}{6}}.\ee
The first question we face is: is there a value
$\Lambda$, for which the noncommutative constraints
on the three gauge couplings are met at
$E=\Lambda$ with the experimental initial conditions
\cite{data},
$g_3=1.207 ,\ g_2=0.6507,\ g_1=0.3575$ at $E=m_Z$. The
answer is affirmative, $\Lambda= 10^{10}$ GeV,
$g_3(0)=0.65$, $g_2(0)=0.57$ with
$f_4x=69.5$ and $f_4\t y=159.5$. This value is to be
compared to $\Lambda=10^{13}-10^{17}$ GeV,
$g_3(0)=g_2(0)=0.52-0.56$ in the stiff case where the
three gauge coupling constraints cannot be fit by one
single $\Lambda$. 

If we want the perturbative calculation to make
sense, we must require that ---  in the huge energy 
range from 100 MeV, where strong forces supposedly
show confinement, all the way up to the
noncommutative cut off $\Lambda$ --- all couplings
remain positive and all dimensionless couplings must
remain smaller than unity. In the following,
we neglect all fermion masses with respect to the top
mass so that we are left with three additional
couplings, the Higgs selfcoupling
$\lambda$, the Yukawa coupling Higgs-top
$g_t=m_t/v$ and the only dimensional coupling
$\mu^2$. Their
$\beta$ functions are \cite{jones},
\bb \beta_\lambda&=& {\textstyle\frac{1}{16\pi^2}}
(96\lambda^2+24\lambda g_t^{2}-6g_t^4
-9\lambda g_2^2-3\lambda g_1^2+
{\textstyle\frac{9}{32}}g_2^4+
{\textstyle\frac{3}{32}}g_1^4+
{\textstyle\frac{3}{16}}g_2^2g_1^2),\\
\beta_t&=&{\textstyle\frac{1}{16\pi^2}}
(9g_t^3-8g_3^2g_t-
{\textstyle\frac{9}{4}}g_2^2g_t-
{\textstyle\frac{17}{12}}g_1^2g_t),\\
\beta_{\mu^2}&=&{\textstyle\frac{1}{16\pi^2}}
\mu^2(48 \lambda+12g_t^2-
{\textstyle\frac{9}{2}}g_2^2
-{\textstyle\frac{3}{2}}g_1^2).\ee
Note that the evolution of $\mu^2$ does not take into
account its quadratic divergences. They are
important if $\Lambda$ is considered a cut off with
physical significance as in noncommutative
geometry. They give rise to the fine tuning problem
which, as advocated by Chamseddine \& Connes, solves
the problem from the stiff constraints
(\ref{g23},\ref{stiffl},\ref{stiffm}),
$m_W(\Lambda)=\Lambda /(16\sqrt 2)$. In the soft
action, the constraint on $\mu^2$ disappears.
Furthermore, $\mu^2$ decouples from the other
couplings and 
has no bearing on the validity range of perturbation
theory. 

We integrate the differential equations for
$\lambda(t)$ and $g_t(t)$ numerically with `initial
values' 
\bb g_t(\log m_Z/\Lambda)&=&\frac{m_t}{v}\,=
{\textstyle\frac{1}{2}}g_2(\log m_Z/\Lambda)
\,\frac{m_t}{m_W}\,,\\ 
\lambda(0)&=&\frac{1}{3}g_3^2(0).\ee 
from the
noncommutative constraint (\ref{softlam}). With 
$m_W=80$ GeV and $m_t=180\pm12$ GeV we get
the Higgs mass (at the pole)
\bb m_H=4\sqrt2\,\frac{\sqrt{\lambda(\log
m_Z/\Lambda)}}{g_2(\log m_Z/\Lambda)}\,m_W\,=\,
235\pm3\ {\rm GeV}.\ee
Note that in presence of the noncommutative
constraints (\ref{softb}) and (\ref{softlam}),
$\lambda={\textstyle\frac{1}{3}}g^2_3$, the
hypothesis of the big desert implies the
validity of perturbation theory throughout this
desert.

\section{Connes-Lott versus Chamseddine-Connes}

In flat spacetime, the Connes-Lott (CL) and the
Chamseddine-Connes (CC) actions coincide up to a
possible cosmological constant. While the constraints
on the gauge group and on the fermion and Higgs
representations are identical in both approaches, this
is not so for the constraints on the coupling constants.
Let us recall the constraints on the couplings from 
Connes-Lott \cite{ks}, 
\bb g^{-2}_3&=&{\textstyle\frac{4}{3}}N\tilde
x,\label{g3}\\
 g_2^{-2}&=&{Nx+y_1+y_2+y_3},\label{g2}\\
g^{-2}_1&=&Nx+{\textstyle\frac{2}{9}}N\tilde
x+{\textstyle\frac{1}{2}}(y_1+y_2+y_3)
+{\textstyle\frac{1}{2}}N\tilde
y,\label{g1soft}\\
\lambda&=&{\textstyle\frac{1}{16}}K/L^2,\\
&&L\ =\ x\t M_u^*M_u+x\t M_d^*M_d+\t y
M_e^*M_e,\\
&&K\ =\ {\textstyle\frac{3}{2}} x\t
\left[M_u^*M_u\right]^2+
{\textstyle\frac{3}{2}}x\t
\left[M_d^*M_d\right]^2+{\textstyle\frac{3}{2}}\t y
\left[M_e^*M_e\right]^2\cr &&\qq\qq\qq\qq
+x\t \left[M_uM_u^*M_dM_d^*\right]
+{\textstyle\frac{1}{2}}\left[
\frac{1}{Nx+\t y}\,+\,\frac{1}{Nx+\t y/2+N\tilde y/2}
\right],\\
\mu^2&=&K/L,\\
\Lambda_C&=&0.\ee
The differences have two origins. The first is: while
the CL action is the square of a 2-form that
brings in its junk, the CC action is
computed from 1-forms only, the connections in the
covariant Dirac operator. 1-forms do not carry junk. 
 This accounts for the difference between $L_2$ and
$K$ and for the different Higgs masses $\sqrt 2\mu$.
The second origin is more conceptual. The CL
scheme affords the possibility to use two Dirac
operators, the usual covariant one in the Dirac action
for the fermions, and another one to build the bosonic
action. This other one is covariant as well, but its
couplings between fermions and gauge bosons break
charge conjugation. The bosonic part of the CL action
is a Dixmier trace over all fermions and anti-fermions
of the square of the curvature. This trace of course
leaves no trace of charge conjugation violation but
allows for more flexible gauge couplings, i.e. two
additional positive parameters, $\tilde x$ and $\tilde
y$. As explained in the introduction, Connes' extended
principle of relativity forbids the use of the charge
conjugation violating Dirac operator. One may of
course choose to use only the symmetric Dirac
operator also in the CL approach. Then, the
constraints on the gauge couplings become as in CC
\cite{russ}.

What are the physical consequences of the
constraints? From the CC action we get
$\frac{1}{4}<\sin^2\theta_w<\frac{9}{20}$ whereas
today experimentally $\sin^2\theta_w=0.2315\pm
0.0005$ at the $Z$ mass forcing upon us a cut off
$\Lambda$ of at least $10^{10}$ GeV and the big desert.
Trusting perturbative quantum field theory
extrapolated all the way up to science fiction energies
we get a Higgs mass (at the pole) $m_H=235\pm3\ {\rm
GeV}$ for
$\Lambda= 10^{10}$ GeV, $m_H=221\pm5\ {\rm GeV}$
for
$\Lambda=10^{17}$ GeV.

 The CL action on the other hand implies
$0<\sin^2\theta_w<\frac{8}{15}$ and can well live
without big quantum corrections, the Higgs mass is
determined essentially by the top mass, $m_H = 298 
\pm 21 {\rm \, GeV}$   if $m_t  = 180 \pm12
{\rm \, GeV}$. The cut off $\Lambda$ can be taken
of the order of the Higgs or top mass where it should
be visible experimentally soon.

\section{Outlook}

We conclude with a list of open questions and wishful
thinking.
\begin{itemize}
\item
Let $\aa$ be the associative algebra ${\cal
C}^\infty(M)\ot(\hhh\op\cc\op M_3(\cc))$. This is
the algebra of the internal space in the standard
model.
$\aa$ fits well with experiment, but $\aa$ is ugly.
\item
The basic variable is the Dirac operator acting on
fermions. The fermions must define a representation of
an associative algebra and are constrained by the
axioms of noncommutative geometry, i.e. of spectral
triples \cite{congr}. These axioms still leave many
choices \cite{kraj}, one of which the quarks and
leptons of the standard model with their mass matrix
taken from experiment. Of course, we want an
explanation for this choice. We have softened the
action by allowing arbitrary positive parameters,
$f_0,\ f_2,\ f_4$ from the function $f$ and $x,\ y_1,\
y_2,\ y_3$ from the element $z$ of the commutant.
These parameters poise the trace just as gauge
couplings in a Yang-Mills theory. We also want an
explanation for these numbers.
\item
To define the Dirac operator in Riemannian geometry,
the spin group is essential. There is no generalization
of the spin group to noncommutative geometry yet.
According to Connes \cite{congr}, this should be a
quantum group and it should help us to get a handle on
the arbitrary choices above \cite{qg}.
\item
Noncommutative geometry grew out of quantum
mechanics. Noncommutative geometry unifies gravity
with the subnuclear forces. We expect noncommutative
geometry to reconcile gravity with quantum field
theory and perhaps at the same time Connes-Lott and
Chamseddine-Connes.
\item
Minkowskian geometry explains the magnetic field,
Riemannian geometry explains gravity. Both
geometries have operated revolutions on spacetime
that today are well established  experimentally: the
loss of absolute time and the loss of universal time.
How can we observe the noncommutative nature
 of time, its uncernainty or `fuzziness', and what is its
characteristic scale
$1/\Lambda$? A hand waving argument combining the
Schwarzschild radius with Heisenberg's uncertainty
relation indicates
$\Lambda<m_P$. 
\item
So far noncommutative geometry is developed in
Euclidean, compact spacetimes, so `Wick rotation' and
3+1 split remain to be understood \cite{kal}.
After this, we expect noncommutative geometry to
change our picture of black holes in a similar fashion
that Heisenberg's uncertainty relation has cured the
Coulomb singularity of the hydrogen atom. Also our
picture of the big bang, cosmology and the origin of
time is expected to be revised \cite{rov}.
\end{itemize}
Planatary motion has degraded circles to epicycles and
dismissed them all together in favour of ellipses.
Particle physics is about to dismiss Riemannian
geometry in favour of noncommutative geometry and
the question is, what dynamics is behind these new
ellipses?

 \end{document}